\documentclass[preprintnumbers, prd, twocolumn, showpacs, floatfix, preprintnumbers, letterpaper, superscriptaddress,nofootinbib]{revtex4}
\usepackage{amsfonts}
\usepackage{eurosym}
\usepackage[dvips]{graphicx}
\usepackage{epsf}
\usepackage{amsmath}
\usepackage{amssymb}
\usepackage{graphicx}
\usepackage{dcolumn}
\usepackage{bm}
\usepackage{color,wasysym}
\usepackage{amsmath,amsfonts,amssymb}
\usepackage{latexsym}
\usepackage{dcolumn}
\usepackage{graphicx,epsfig}
\usepackage{amsthm}
\usepackage{color}
\usepackage{hyperref}
\begin{document}

\title{Emergent Cosmos in  Einstein-Cartan Theory}
\author{H. Hadi}
\affiliation{Department of Physics, Azarbaijan Shahid
Madani University, Tabriz,
53714-161,
Iran}
\author{ Y. Heydarzade}
\email{heydarzade@azaruniv.edu; Corresponding author} 
\affiliation{Department of Physics, Azarbaijan Shahid
Madani University, Tabriz,
53714-161,
Iran}
\affiliation{ Research Institute for Astronomy and Astrophysics of Maragha (RIAAM),
Maragha
55134-441, Iran}
\author{M. Hashemi}
\email{m\_hashemi@sbu.ac.ir} \affiliation{Department of Physics, Shahid Beheshti University, G. C., Evin,Tehran, 19839, Iran}
\author{F. Darabi}
\email{f.darabi@azaruniv.edu}
\affiliation{Department of Physics, Azarbaijan Shahid
Madani University, Tabriz,
53714-161,
Iran}
\affiliation{ Research Institute for Astronomy and Astrophysics of Maragha (RIAAM),
Maragha
55134-441, Iran}
\begin{abstract}
 Based on the Padmanabhan's proposal, the accelerated expansion of the universe
can be driven by the difference between the
surface and  bulk degrees of freedom   in a
region of space, described    by the relation
 $ dV/dt = N_{sur}-N_{bulk}$  where $N_{sur}$ and $N_{bulk}=  -N_{em}  +N_{de}$ are  the degrees of freedom assigned
to the surface area and the matter-energy content inside the bulk such that the indexes $``em"$ and $``de"$  represent energy-momentum and dark energy, respectively. In the present work,  the
dynamical effect of the Weyssenhoff perfect fluid with intrinsic spin and its corresponding spin degrees of freedom 
in the framework
of Einstein-Cartan (EC) theory are investigated.  Based on
the modification of Friedmann equations due to the spin-spin interactions,
a correction term for the Padmanabhan's
original relation $dV /d t=N_{sur}+N_{em}  -N_{de}$ including the number of degrees of freedom related to this spin interactions is obtained through the modification
in $N_{bulk}$ term
as $N_{bulk}=  -N_{em}+N_{spin} +N_{de}$ leading
to $dV /d t=N_{sur}+N_{em}-N_{spin} -N_{de}$ in which $N_{spin}$ is the corresponding degrees of freedom related
to the intrinsic spin of the matter content of the universe. 
Moreover, the validity of the unified first law and the generalized second law of thermodynamics for the Einstein-Cartan cosmos are investigated. Finally, by considering the covariant entropy conjecture and the bound resulting
from the emergent scenario, a total entropy bound is obtained.
Using this bound, it is shown that the for the universe as an expanding thermodynamical system, the total effective Komar energy never exceeds  the square of the expansion rate with a factor of $\frac{3}{4\pi}$.
\end{abstract}

\maketitle

\section{Introduction}
 According to the current researches, one can obtain the gravitational field equations in the same way that the equations of an emergent phenomena like
fluid mechanics or elasticity are derived \cite{Paddy:2010nnn, universe:1968, jacobson:1995mm}. 
In the framework of the emergent gravity model, the Einstein gravitational field equations can be derived
from the thermodynamics principles with some extra assumptions
\cite{Paddy:2010nnn, Paddy:2010ljoi}.  Therefore, Einstein field equations can be understood as spacetime equations of state \cite{jacobson:1995mm}. By assuming the existence
of a spacetime manifold, its metric and  curvature, Padmanabhan has treated the  Einstein field equations as  an emergent phenomenon  \cite{Paddy:2012kshj}.
It has been proposed that in a cosmological context, the accelerated expansion of the universe \cite{Riess:1998}
can be obtained from the difference between the surface  and 
bulk degrees of freedom denoted by the relation
$\Delta V/\Delta t = N_{sur}-N_{bulk}$ where $N_{sur}$ and $N_{bulk}$ are  the corresponding  degrees of freedom related
to the surface area, matter-energy content (or dark matter (DM) and dark energy (DE)) inside the bulk space, respectively \cite{Paddy:2012nnc}. Different
cosmological models have been proposed to explain the late time accelerated expansion of the universe \cite{Riess:1998}.
One of these cosmological models\ is known as the  dark energy model where the universe is supposed to be dominated by a dark fluid possessing a negative pressure \cite{Alam:2004,Ratra:1988,Armendariz:2001} (for a review, see \cite{Copeland:2006}).  Violation of the strong energy condition is a feature of this dark fluid, i.e $\rho+3p>0$. On the other hand, the modified gravity theories, such as $f(R)$ gravity \cite{Felice:2010}, $f(T)$ gravity  
\cite{Linder:2010}, Weyl gravity
\cite{Mannheim:1989}, Gauss-Bonnet gravity
\cite{Wheeler:1986}, Lovelock gravity \cite{Lovelock:1971}, Ho\v{r}ava-Lifshitz gravity  
\cite{Horava:2009}, massive gravity \cite{Rham:2011}, heterotic string theory \cite{string:1993} and braneworld scenarios \cite{brane:2002}, are another approaches for explaining the late time accelerated expansion of the universe. In these modified models,
the additional terms  in the gravitational Lagrangian play the role of an effective dark energy component with a geometric origin rather than an {
\it ad hoc} introduction of the dark energy sector with unusual physical features.    These cosmological
models explaining the current accelerated expansion phase possess a series of conditions and constraints arising
from various laws of physics such as thermodynamics laws \cite{thermo:2001} or astrophysical
data.
 In this way, four laws of black hole mechanics driven from the classical Einstein field equations are implemented to explain  the structure of spacetime and its relation with thermodynamical behaviour of the system \cite{thermo:1973, thermo:1975}. In the significant pioneering research, Jacobson proved that the classical  general relativity (GR) behaves like thermodynamical system (for example, surface gravity could be understood similar to temperature in thermodynamical system) \cite{thermo:1995}. Then, the Einstein field equations were obtained from the relation of entropy and horizon area together with   the Clausius relation $dQ= TdS$ where $Q$, $S$ and $T$ are the heat, the entropy and the temperature,
respectively.  In this regard, where the connection between gravity and thermodynamics holds, the Friedmann equations are obtained by applying the first law of thermodynamics to the apparent horizon of the  FLRW universe \cite{fl:2007, fl:2007hh, fl:2005, fl:2003, fl:2000, fl:2007dc, fl:2007sr, fl:2077kih, Paddy:2010bbs, Hashemi:2015, Hashemi:2015mm}.
The  second law of thermodynamics and its generalized version
is also studied in different modified gravity models such as \cite{Sharif:2013, sl:2013,
sl:2006km,sl:2077de,sl:2010mwjh,sl:2008, sl:2099,sl:2006fyf,sl:2010viuf,sl:2004dfgf}. 

On the other hand, a cosmological model
is  influenced by  the choice of a matter field source which is coupled with the Einstein equations through its energy-momentum tensor. Usually, the matter source of the universe is considered as a
perfect fluid or scalar fields  \cite{Scalarfield:2006}.  Regarding the early stage of the Universe when its matter content possesses an additional intrinsic spin property, it is necessary to consider
a classical spin fluid or even a massless or massive spinor fields
 as the matter source \cite{spinorfield:2003}. In 1923,  \'{E}lie Cartan introduced a modification of the Einstein general theory of relativity (GR) which nowadays
is known as Einstein-Cartan (EC) theory \cite{cartan:1923, cartan:2010}. In this framework,
 a relation between the intrinsic angular
momentum of matter source and the  spacetime torsion is introduced before introducing this intrinsic angular
momentum as the spin into quantum theory by
Goudsmit and Uhlenbeck in 1925. The classical spin can be
introduced in general relativity  in two distinct ways. As the first method, one can consider spin  as a dynamical quantity without changing the Riemannian structure of the geometry of the background  spacetime \cite{spin:1937}. The
spin introduced in this way is similar
to the spin of quantum mechanics ~and the Dirac theory of
the electron.  In the second approach, as
introduced by Cartan,  the structure of
 spacetime is generalized to possess
torsion as well as curvature by considering the metric and the non-symmetric affine connection as independent quantities \cite{Wat:2004}.    This
Riemann-Cartan geometry is usually denoted by $U_4$ in order to distinguish it from the Riemannian geometry.  After Cartan's  research, many other efforts have been made by Hehl \cite{Hehl:1976}, Trautman \cite{Trautman:1973}
and Kopczynski \cite{Kop:1972} to bring spin into the curved spacetime. This approach allowed  one to define the torsion of spacetime 
and its connection  with spin.  In the context of EC theory, torsion does
not appear as a dynamical quantity rather it can be represented  in
terms of the spin sources by matter fields with
intrinsic angular momentum \cite{Hehl:1976}.  Most of the     researches on the cosmological applications of the EC theory have been made with the semiclassical  spin fluid possessing the energy density $\rho_s$, pressure $p_s$ and spin density vector $S^\alpha$   which is orthogonal to  the four velocity vector $u^\alpha$  of the spin fluid in the comoving reference frame of the fluid.
This generalization of the perfect fluid with spin is known as the 
Weyssenhoff  fluid where its dynamics   was  comprehensively studied by Weyssenhoff
and other researchers \cite{Weys:1947}. 
Similar to the other alternative theories of gravity,  the
cosmological solutions of the EC theory possessing the spin matter source
and their influence on
the  structure and dynamics of the universe are extensively investigated. These studies include  the effects of
torsion and spinning matter in a cosmological setup and its
possible role to solve the   singularity problems, pre-Friedmann stages of evolution,  inflationary expansion,
the late time accelerated expansion of the Universe, rotation of the Universe
and gravitational collapse and so on \cite{Kuch:1978}.

In this paper, we investigate the emergent universe
scenario and its thermodynamical aspects in the framework of EC theory. By considering the modifications to Friedmann equations of the EC theory, we discuss on Padmanabhan's relation and thermodynamical features of the
model. This paper is organized as follows. In section II, we review the EC theory. In section III, we study the issue of emergence of  spacetime in the context of this model. In section IV, thermodynamics of  the Einstein-Cartan universe is investigated. {In section V, we discuss on the Covariant Entropy Conjecture and Emergent Universe scenario in Einstein-Cartan
theory.} Finally,  in the last section, our concluding remarks are represented. Also, we consider the units of $c=1$ with   metric signature $(+,-,-,-)$ of  spacetime. Also, we use the signs [] and () for denoting antisymmetric and  symmetric parts, respectively.

\section{The Einstein-Cartan Model}
The Einstein-Cartan theory can be driven using the following action
\begin{equation}\label{1}
{\cal S}=\frac{1}{16\pi G}\int d^4x \sqrt{-g}\tilde R+\int d^4x\sqrt{-g}{\cal L}_M,
\end{equation}
where $\tilde R$ and $\mathcal{L}_{m}$ are the Ricci scalar associated to the asymmetric connection $\tilde\Gamma^{}_{}$ and the Lagrangian density of matter fields coupled to the gravity, respectively. 

 The asymmetric connection $\tilde\Gamma^\mu_{\,\,\,\,\alpha\beta}$  can be written in terms of the Levi-Civita connection $\Gamma^\mu_{\,\alpha\beta}$ as
\begin{eqnarray}\label{2}
\tilde\Gamma^\mu_{\,\,\,\,\alpha\beta}=\Gamma^\mu_{\hspace{.2cm}\alpha\beta}+K^\mu_{\hspace{.2cm}\alpha\beta},
\end{eqnarray}
where $K^\mu_{\hspace{.2cm}\alpha\beta}$,  known as the 
``contorsion tensor", which  is related to the torsion
($Q_{\alpha\beta}^{\hspace{.5cm}\mu}:=\tilde
\Gamma_{[\alpha\beta]}^{\hspace{.5cm}\mu}$ ) as \cite{Hehl:1976}
\begin{eqnarray}\label{3}
K^\mu_{\hspace{.2cm}\alpha\beta}:
=\frac{1}{2}\left(Q_{\hspace{.2cm}\alpha\beta}^{\mu}-Q^{\hspace{.3cm}\mu}_{\alpha\hspace{.4cm}\beta}-Q^{\hspace{.3cm}\mu}_{\beta\hspace{.4cm}\alpha}\right).
\end{eqnarray}
Using the variation of the action with respect to the metric $g_{\mu\nu}$ and contorsion tensor
$K^{\mu}_{\alpha\beta}$ , one can find the dynamical
equations of motion of the theory as follows \cite{Hehl:1976, gasp}
\begin{eqnarray}\label{4}
G^{\mu\nu}&-&\left(\tilde \nabla_{\alpha}+2Q_{\alpha\beta}^{\hspace{.4cm}\beta}\right)\left(T^{\mu\nu\alpha}-T^{\nu\alpha\mu}+T^{\alpha\mu\nu}\right)\nonumber\\
&=&8\pi GT^{\mu\nu},\nonumber\\
T^{\mu\nu\alpha}&=&8\pi G\tau^{\mu\nu\alpha},
\end{eqnarray}
where $G^{\mu\nu}$ and $\tilde \nabla_\alpha$ are  the Einstein tensor and covariant derivative  based on the asymmetric connection $\tilde \Gamma^{\mu}_{\alpha\beta}$
, respectively, and $T_{\mu\nu}^{\hspace{.3cm}\alpha}$ is defined in terms of the torsion
tensor ${Q_{\mu\nu}}^{\alpha}$ as
\begin{equation}\label{5}
T_{\mu\nu}^{\hspace{.3cm}\alpha}=Q_{\mu\nu}^{\hspace{.3cm}\alpha}+\delta_\mu^\alpha
Q_{\nu\beta}^{\hspace{.4cm}\beta}-\delta_\nu^\alpha
Q_{\mu\beta}^{\hspace{.4cm}\beta}.
\end{equation}
We also  defined
\begin{eqnarray}\label{6}
&&T^{\mu\nu}:=\frac{2}{\sqrt{-g}}\frac{\delta{\cal L}_M}{\delta g_{\mu\nu}},\nonumber\\
&&\tau^{\mu\nu\alpha}:=\frac{1}{\sqrt{-g}}\frac{\delta{\cal
L}_M}{\delta K_{\alpha\nu\mu}},
\end{eqnarray}
as the energy-momentum tensor and the spin 
 density tensor,
respectively. Combining equations  (\ref{4}) and (\ref{5}),  one can obtain the Einstein field equations  with a modification in the energy-momentum as
\begin{equation}\label{E}
G^{\mu\nu}=8\pi G(T^{\mu\nu}+\tau^{\mu\nu}),
\end{equation}
where
\begin{eqnarray}\label{8}
\tau^{\mu\nu}&=&8\pi G\{
-4\tau^{\mu\alpha}_{\hspace{.4cm}[\beta}\tau^{\nu\beta}_
{\hspace{.3cm}\alpha]}-2\tau^{\mu\alpha\beta}\tau^\nu_{\hspace{.3cm}\alpha\beta}
+\tau^{\alpha\beta\mu}\tau_{\alpha\beta}^{\hspace{.3cm}\nu}\nonumber\\
&&+\frac{1}{2}g^{\mu\nu}
\left(4\tau_{\lambda
\hspace{.1cm}[\beta}^{\hspace{.2cm}\alpha}\tau^{\lambda\beta}_{\hspace{.4cm}\alpha]}+
\tau^{\alpha\beta\lambda}\tau_{\alpha\beta\lambda}\right)\},
\end{eqnarray}
is the correction term for the energy-momentum tensor generated by the spacetime
torsion
\cite{Weys:1947}. If the torsion, or spin, vanishes then $\tau ^{\alpha \beta}$  vanishes  and
 the standard Einstein field equations ( $G^{\mu\nu}=8\pi GT^{\mu\nu}$) are recovered. Suppose
that the Lagrangian ${\cal L}_M$
represents a fluid of spinning particles in the early Universe
minimally coupled to the metric and the torsion of the $U_4$ theory.
In this context, one can employ a classical description of spin as postulated by Weyssenhoff and Raabe, which is given by \cite{Weys:1947}
\begin{equation}\label{FC}
\tau_{\mu\nu}^{~~\alpha}=\frac{1}{2}<S_{\mu\nu}>{ u}^{\alpha},~~~~~~<S_{\mu\nu}{ u}^{\mu}>=0,
\end{equation}
where ${ u}^{\alpha}$ is the four-velocity of the fluid element and $S_{\mu\nu}=-S_{\nu\mu}$ is a second-rank antisymmetric tensor which is defined as the spin density tensor. The spatial components of this tensor include the 3-vector $(S^{23},S^{13},S^{12})$ which coincides in the rest frame with the spatial spin density of an element of the matter fluid. The remaining spacetime components i.e., $(S^{01}, S^{02}, S^{03})$ are assumed to be zero in the rest frame of the fluid element. Such an assumption can be covariantly formulated as the constraint given in the second part of (\ref{FC}). This constraint on the spin density tensor is usually called the Frenkel condition which requires the intrinsic spin of matter to be spacelike in the rest frame of the fluid. More precisely, this condition leads to an algebraic
relation between the spin density and torsion tensor as 
\begin{equation}T_{\nu}={T^{\mu}}_{\nu\mu}= <u^{\mu}S_{\nu\mu}>,
\end{equation}
 which can also be recovered directly from the formalism
proposed in \cite{1}. Therefore, the Frenkel condition implies that the only
remaining degrees of freedom of the spacetime torsion are its traceless components. The spinning fluid (fluid that possesses an internal angular momentum density) introduced in this way is called the \lq\lq{}Weyssenhoff fluid\rq\rq{}, which generalizes the perfect fluid of general relativity to the case of non-vanishing spin. The Weyssenhoff fluid is a continuous medium that the elements of which are characterized (together with the energy and momentum) by the intrinsic angular momentum (spin) of its constituent particles, see also \cite{2} and \cite{3}. 

The energy-momentum tensor can be decomposed as 
\begin{equation}\label{gogoli}
T^{\mu\nu}
= T_P^{\hspace{.1cm}\mu\nu}+
T_S^{\hspace{.1cm}\mu\nu},
\end{equation}
where 
$T_P^{\hspace{.1cm}\mu\nu}$ and 
$T_S^{\hspace{.1cm}\mu\nu}$are the usual perfect fluid and the intrinsic-spin fluid part as 
\begin{eqnarray}\label{10}
&T_P^{\hspace{.1cm}\mu\nu}&=(\rho+p)u^{\mu}u^{\nu}-pg^{\mu\nu},\nonumber\\
&T_S^{\hspace{.2cm}\mu\nu}&=u^{(\mu}S^{\nu)\alpha}u^\beta
u_{\alpha;\beta}+\nabla_{\alpha} (u^{(\mu}S^{\nu)\alpha})\nonumber\\
&&\,\,\,\,\,\,\,+Q_{\alpha\beta}^{\hspace{.2cm}(\mu}u^{\nu)}
S^{\beta\alpha}-u^\beta
S^{\alpha(\nu}Q^{\mu)}_{\hspace{.1cm}\alpha\beta} \nonumber\\
&&\,\,\,\,\,\,\,-\omega^{\alpha(\mu}S^{\nu)}_{\hspace{.3cm}\alpha}+u^{(\mu}S^{\nu)\alpha}\omega_{\alpha\beta}u^\beta,
\end{eqnarray}
$\omega$ being the angular velocity corresponding to the intrinsic spin, and $\nabla_\mu$ represents the covariant derivative associated to the symmetric Levi-Civita connection $\Gamma^\mu_{\,\alpha\beta}$.

From the microscopic viewpoint, a randomly oriented gas of fermions is the source for spacetime torsion. However, we have to deal with this issue at a macroscopic level, that is, we need to perform suitable spacetime averaging. In this sense, the average of spin density tensor vanishes, $<S_{\mu\nu}>=0$ \cite{{Hehl:1976,GasPRL}}. However, though the vanishing of this term macroscopically, the square of spin density tensor $< S_{\mu\nu}S^{\mu\nu}>=2\sigma^2$ would  make its own contribution to the total energy momentum tensor, in such a way that the field equations in Einstein-Cartan theory are different
 from those in general relativity even in the classical macroscopic limit \cite{{Hehl:1976,GasPRL}}. 
 A suitable averaging process then gives (see the Appendix A)
 
\begin{eqnarray}\label{12}
&&<\tau^{\mu\nu}>=4\pi G \sigma^2u^\mu u^\nu +2\pi G
\sigma^2g^{\mu\nu},\nonumber\\
&&<T_S^{\hspace{.2cm}\mu\nu}>=-8\pi G\sigma^2u^\mu u^\nu.
\end{eqnarray}
Indeed, since the right hand side of (\ref{E}) includes  the $\tau^{\mu\nu}$
contribution with a quantum origin, the quantities in the right hand of (\ref{8})
must be replaced
by their expectation values. For more detail on the conditions under  which  the  expectation  value  of the energy-momentum tensor can act as the source for a semiclassical gravitational field, see  the works in \cite{semi}.  Therefore, the Einstein field  equations (\ref{E}) read as 
\begin{equation}\label{G}
G^{\mu\nu}=8\pi G\Theta^{\mu\nu},
\end{equation}
where $\Theta^{\mu\nu}$ represents the effective macroscopic
limit of matter fields defined as
\begin{eqnarray}\label{14}
\Theta^{\mu\nu}&:=&<T^{\mu\nu}>+<\tau^{\mu\nu}>\nonumber\\
&=&(\rho+p-\rho_{s}-p_s)u^\mu u^\nu-\left(p-p_{s}\right)g^{\mu\nu}\nonumber\\
&=&\left(\rho+p-4\pi
G\sigma^2\right)u^\mu u^\nu\nonumber\\
&&-\left(p-2\pi
G\sigma^2 \right)g^{\mu\nu}.
\end{eqnarray}
Then, one may consider the following forms for the total energy density
and pressure which support the field equations 
\begin{equation}\label{rev1}
\rho_{tot}=\rho-\rho_{s},\hspace{5mm}p_{tot}=p-p_{{s}}, 
\end{equation} 
where $\rho_{s}=p_{s}=2\pi G \sigma^2$. From this, it is seen that $p_{s}/\rho_{s}=1$ and consequently  the spin matter behaves as a fictitious fluid with an
equation like that of the Zeldovich stiff matter. 
Beside the works which assume a classical form of the spin fluid as the source of torsion, it is worth mentioning that  a full quantum treatment has been recently done in \cite{Lucat:2015rla}.

\section{Emergence of Spacetime in Einstein-Cartan Theory}
We consider a homogeneous and isotropic
universe described by the  FLRW metric
\begin{equation}
ds^2 =dt^{2}-a(t)^2 (\frac{dr^2}{1-kr^2}+r^2(d\theta^2 + sin^2\theta d\phi^2)).
\end{equation}
where $k=0,\pm1$ represents spatial curvature of the universe
(in the following we will focus on the flat universe).
Then, using
equations (\ref{G}) and (\ref{14}), Friedmann equations will be
\begin{equation}\label{a}
H^{2}=\frac{8\pi G}{3}(\rho-2\pi G\sigma^{2})=\frac{8\pi}{3}\rho_{tot},
\end{equation}
 \begin{equation}\label{b}
\dot H+H^{2}=\frac{-4\pi G}{3}(\rho+3p-8\pi\sigma^{2})=\frac{-4\pi
G}{3}(\rho+3p)_{tot}.
 \end{equation}
We note that, $\rho_{total}=\rho-\rho_s$ is assumed to be positive. In this respect, from one hand, the spin squared term is proportional to $a^{-6}$ and from the other its coupling constant is proportional to the square of gravitational coupling constant which makes the spin effects to be crucial only at extremely high matter densities, e.g. at the late stages of collapse scenario or in the early times of the evolution of the universe. Such effects could provide non-singular cosmological \cite{Trautman:1973,4}  as well as astrophysical settings \cite{jalal2015}. However, during such extreme regimes, though the weak (or null) energy condition may be violated due to the negative pressure contribution due to the spin contribution, the total energy density as given in equation (\ref{a}) remains positive; the matter density $\rho$ could be proportional to $a^{-6}$ for a stiff fluid and in competition to the spin density, these two terms at the right hand side of equation (\ref{a}) could be at most equal leading to a vanishing Hubble parameter. For more
discussion on the negative contribution of spin fluid, see the Appendix B.
On the other hand, the combination of equations (\ref{a}) and (\ref{b}) gives the following generalization of the covariant energy conservation law including the spin term 
\begin{equation}\label{conservation}
\frac{d}{dt}(\rho-2\pi G\sigma^{2})=-3H(\rho+p-4\pi G\sigma^{2}),
\end{equation}
where we can consider the filling matter field as an unpolarized fermionic perfect fluid with the barotropic equation of state $p=\omega \rho$. By decomposition
of the matter source in equation (\ref{gogoli}), we can treat the above conservation
law for two non-interacting fluids. Therefore, it gives
\begin{equation}\label{magol}
\rho=\rho_{0}a^{-3(1+\omega)},
\end{equation}
where $\rho_0$ is energy density at present time. 
Eq. ({\ref{conservation}}) could be rearranged in the following form
\begin{eqnarray}\label{conservation-re}
\frac{d}{dt}\rho+3H(\rho+p) &=& 2\pi G(\frac{d}{dt}\sigma^{2}+6H\sigma^{2})\nonumber\\
&=&\frac{2\pi G}{\sigma^{2}}\frac{d}{dt} \ln (\sigma^{2}a^6).
\end{eqnarray}
By implementing Eq.(\ref{magol}) the LHS equals to zero. By reusing Eq.(\ref{magol}),
 one could reach
\begin{equation}\label{sigma}
\sigma^{2}=\frac{\hslash}{8}(\frac{\rho_0}{A_{\omega}})^{\frac{2}{1+\omega}} a^{-6}=\frac{\hslash}{8}(\frac{\rho}{A_{\omega}})^{\frac{2}{1+\omega}},
\end{equation}
and
\begin{equation}\label{rhos}
{\rho_{s}=2\pi G\sigma^2 =\rho_{0s}a^{-6}},
\end{equation}
in which $A_{\omega}$ is a dimensional constant depending on $\omega$ and
$\rho_{0s}=\frac{\hbar}{8}A_{\omega}^{\frac{-2}{1+\omega}}\rho_{0}^{\frac{2}{1+\omega}}$
\cite{Nur:1983}.

Multiplying equation (\ref{b})
 by $-4\pi H^{-4}$, we get
 \begin{equation}
-4\pi \frac{\dot H}{H^{4}}= \frac{4\pi}{H^{2}}+ \frac{16 \pi^{2}G(\rho + 3p)_{tot}}{3H^{4}}.
 \end{equation}
 Assuming $V=4 \pi H^{-3}/3$ as the spherical volume
with the Hubble radius $H^{-1}$, namely the Hubble volume, we have
\begin{equation}\label{dVt1}
\frac{dV}{dt}=-4\pi\frac{\dot H}{H^4}=\frac{4\pi}{H^2}
+\frac{16\pi^{2}G(\rho + 3p)_{tot}}{3H^{4}}.
 \end{equation}
On the other hand, according to Padmanabhan's idea, the number of degrees of freedom on the spherical surface of the Hubble radius $H^{-1}$ is
given by \cite{Paddy:2012nnc}
\begin{equation}\label{sur}
N_{\mathrm{sur}}=\frac{A}{L_{P}^2}=\frac{4\pi
}{L_{P}^2H^{2}},
\end{equation}
where $L_{p}$ is the Planck length and $A=4\pi H^{-2}$ represents the
area of the Hubble horizon\footnote{Note that the area law $S=A/4{L_{p}^{2}}$  as the saturation of the Bekenstein limit \cite{Bek:1981} is completely justified solely in the context of general relativity and  is not correct in general in modified theories, see \cite{ent}.
However, one may argue that true gravitational degrees
of freedom are that of GR only, and the effect of torsion is to modify the right hand side and effectively acts as an additional energy-momentum tensor, restoring the $A/4{L_{p}^{2}}$ law.
}.
Also, the bulk degrees of freedom obey the equipartition law
of energy
\begin{equation}\label{Nbulk}
N_{\mathrm{bulk}}=\frac{2|E_{tot}|}{k_{B}T},
\end{equation}
where $E_{tot}$, $k_B$ and $T$ are the total energy inside of the bulk, the Boltzmann constant and the temperature of the bulk space, respectively. 
 In the following of the paper, we  use natural unit ($k_{B}=c=G=L_{p}=1 $) for the
sake of simplicity.
We also consider the temperature associated with the Hubble horizon
as the Hawking temperature  $T=H/2\pi $, and the energy contained inside the Hubble volume
in Planck units $V=4\pi /3H^{3}$  as the Komar energy
\begin{equation}\label{Komar}
E_{\mathrm{Komar}}=|(\rho +3p)_{tot}|V.
\end{equation}
Based on the novel idea of Padmanabhan,  the cosmic expansion which is conceptually equivalent to the emergence of space is related to  the
difference between the number of degrees of freedom in the
holographic surface and the ones in the corresponding emerged bulk \cite{Paddy:2012nnc}.
Equations (\ref{Nbulk}) and (\ref{Komar}) with Hawking
temperature  will give the bulk degrees of freedom  as
\begin{equation}
 N_{bulk} = -\epsilon \frac{2(\rho +3p)_{tot}V}{k_BT} 
\label{nep}
\end{equation} 
where $\epsilon = +1$ denotes $(\rho + 3p)_{tot}<0$ and $\epsilon = -1$ if $(\rho + 3p )_{tot} >0$. Based on Padmanabhan's assumption the universe can be divided as  matter component, respecting the strong energy condition  $(\rho + 3p)_{tot}>0$, and dark energy component, violating the strong energy condition $(\rho + 3p)_{tot}<0$. Hence, the bulk degrees of freedom reads as
  {\begin{equation}
N_{bulk}=  N_{de} -N_{m}
\end{equation} }
where the indexes $``m"$ and $``de"$  represent matter and dark energy, respectively. So, we have
  {\begin{equation}\label{Nbulk1}
N_{de}- N_{m}=-\epsilon\frac{16\pi^2}{3}\frac{(\rho+3p)_{tot}}{H^4}.
\end{equation}}
Then, using the equation (\ref{dVt1}), the Padmanabhan's equation can be written as follows
\begin{eqnarray}\label{pad}
\label{dVt} \frac{dV}{dt}&=&N_{sur}-N_{bulk}\nonumber\\
&=&N_{sur}+ N_m-N_{de} ,
 \end{eqnarray}
where $N_{sur}$ and  $N_m$ and $N_{de}$  are given by the equations (\ref{sur}) and (\ref{Nbulk}). {On the other hand, because spin is the degrees of freedom of the matter field filling the universe, regarding the equations
(\ref{rev1}), (\ref{a}) and (\ref{b}), it would be natural to write the total contribution of
matter as $N_\mathrm{m}=N_{em}
-N_{spin}$ where $``em"$ stands for $``energy-momentum"$.  In this regard, one can rewrite the equation (\ref{pad}) as
\begin{equation}\label{pad12}
\frac{dV}{dt}=N_{sur}+N_{em}
-N_{spin} -N_{de} ,
 \end{equation}  where the degrees of freedom related to the spin of matter content are given by}
\begin{equation}\label{spinoo}
N_{spin}=\frac{16\pi V\sigma^{2}}{T}.
\end{equation}
The equation (\ref{pad12}) indicates that there are four modes of degrees of freedom for a cosmos filled by the dark energy fluid and the matter content possessing
spin-spin interactions. {For such a universe, other than the surface degrees of freedom,
the energy-momentum degrees of freedom and the ones related to the dark energy,
there are additional degrees of freedom which lie in its spin sector. In
this line, both of the spin  and dark energy sectors  contribute a ``negative
number of degrees of freedom".
} Moreover, using equations (\ref{spinoo}) and (\ref{rhos}), the spin degrees of freedom will be
\begin{equation}\label{ns}
N_{spin}=\frac{8\rho_{0s}V}{T}a^{-6}.
\end{equation}
 This relation shows that the spin degrees of freedom is vanishing at late time. This is because of that
 the spin density and consequently its contribution to Eq.(\ref{pad}) is  very weak at low energy limits, i.e at the late times of the Universe, in contrast to the high energy limits in the very early universe where  the evolution of universe
 can be considerably affected by  it. 
 On the other hand, although the Universe is not pure de Sitter, however  it evolves
toward  an asymptotically de Sitter phase. Then, in order to reach the  holographic equipartition,  we demand  
${dV}/{dt}\rightarrow0$ in the equation (\ref{pad}) which
requires $N_{sur}=N_{bulk}$. To understand the
 feature of $N_{spin}$, it is better to look at
equation (\ref{pad}) without this term.  Following the discussion of Padmanabhan,
one can consider that $N_{bulk}$  includes two parts. The first  one  is
related to the
normal matter sector respecting the strong energy condition and the second one related to the dark
energy sector violating the strong energy condition \cite{Paddy:2012nnc}. This provides the possibility
of dividing the degrees of freedom
of the bulk into two parts, one arising from the degrees of
freedom of dark energy leading to acceleration and the
other one arising from the degrees of freedom of normal matter leading to deceleration. Then, equation (\ref{pad}), without $N_{spin}$  term, {takes the form of $\frac{dV}{dt}= N_{sur}+N_{em}-N_{de}$.} Therefore,   there is no
hope for reaching the holographic equipartition for a universe without   a dark energy sector \cite{Paddy:2010ljoi}. In reference \cite{yaghoub:2015}, the Padmanabhan's
emergent scenario
is investigated in a general braneworld setup. It is found that the Padmanabhan's
relation takes the form   $ \frac{dV}{dt}=N_{sur}- N_{bulk}- N_{extr}$ where $N_{extr}$ is referred to the degrees of freedom related
to the extrinsic geometry of a four dimensional brane embedded in a higher dimensional ambient
space, while $N_{sur}$ and $N_{bulk}$
are exactly the same as before.
Moreover, it is shown that one can avoid of the  term $N_{de}$ denoting dark energy  which has been previously proposed by Padmanabhan. 
This is because, the geometrical component  $N_{extr}$ arising from the brane extrinsic geometry, representing a geometrical dark energy \cite{Maia:2005}, can play the role of  $N_{de}$.
   However, in
the framework of EC theory, 
the spin term cannot  completely play the role of dark energy or  cosmological constant leading to the satisfaction of holographic equipartition law, because
the corresponding degrees of freedom in equation (\ref{pad})
are vanishing at late time, see equation (\ref{ns}), leading to $\frac{dV}{dt}> 0$ in the absence of dark energy. Then, 
unlike in \cite{yaghoub:2015}, although the spin
sector in EC framework plays an important role in the early stages of universe with a repulsive
gravitational effect, at
late times the cosmological constant or dark energy term proposed by Padmanabhan is required again to achieve the  holographic equipartition in this model.
This fact is in agreement with the result obtained in \cite{Szyd:2004} where the luminosity
distance  is implemented to test the models using the supernovae type Ia observations. There, the  authors showed that although a cosmological model with a spin fluid is admissible
but the cosmological constant is still required to explain the accelerating expansion of the universe.  Consequently, the spin fluid can not be considered as an alternative to the cosmological constant description of the dark energy.
\section{Thermodynamics of an Einstein-Cartan Cosmology}
In recent years,  the connection between gravitation and thermodynamics  have received much attention, see for example \cite{Paddy:2010nnn} and
 \cite{jacobson:1995mm,Jacobson:2015uqa,Chakraborty:2015hna, Mohd:2013zca, Mohd:2013jva}, where the first and second laws of thermodynamics are vastly investigated. Through this section, first we obtain the unified first law of thermodynamics based on the  (0,0) component of the Einstein field equations  introduced by Hayward, see \cite{Hayward:1997jp, Hayward:1998ee} and  \cite{Cai:2006rs}. Then, we investigate the
generalized second law of thermodynamics for the Einstein-Cartan cosmos.
\subsection{Unified First Law of Thermodynamics}
The Hubble horizon $H^{-1}$ can be understood as an apparent horizon of the flat FLRW universe\footnote{The dynamical apparent horizon, i.e. $\tilde{r}_A=a(t)\,r$, can be obtained from the equation $h^{\alpha \beta} \partial_{\alpha} \tilde{r} \partial_{\beta} \tilde{r} =0$ where $h^{\alpha \beta}$ is the non-spherical part of the FLRW metric.}\cite{Cai:2005ra}.
By calculating the derivative of $ \frac{1}{\tilde{r}_A}=H$ with respect to the cosmic time, we easily have $ -d\tilde{r}_A/\tilde{r}_A^3 = \dot{H}H dt$. Also, by implementing the modified Friedmann equation (\ref{a}), we obtain
\begin{equation}\label{}
\frac{-1}{\tilde{r}_A^3}\frac{d\tilde{r}_A}{dt}= \dot{H}H = \frac{4\pi}{3}\frac{d}{dt}(\rho-2\pi\sigma^{2}) = \frac{4\pi}{3}\dot{\rho}_{tot}.
\end{equation}
One can simplify this equation using the generalized conservation law in Eq.(\ref{conservation})  as follows
\begin{equation}\label{c2}
d\tilde{r}_A =4\pi \tilde{r}_A^3 H \left(
\rho_{t}+p_{t} \right) dt,
\end{equation}
where $\rho_{t}$ and $p_{t}$ are total energy density and pressure including both of the normal and spin sectors. The Hawking-Bekenstein entropy is $S=  {A}/4 = \pi \tilde{r}_A^2$. Therefore, using this entropy expression and Eq. (\ref{c2}), we get
\begin{eqnarray}\label{c3}
 \frac{1}{2\pi \tilde{r}_A} dS=d\tilde{r}_A =4\pi \tilde{r}_A^3 H \left(
\rho_{t}+p_{t} \right)
dt.
\end{eqnarray}
On the other hand, the temperature can be obtained as \footnote{The Apparent horizon temperature can be calculated by  $ T_{\mathrm{H}} =\frac{|\kappa|}{2\pi}$ where 
$ \kappa = \frac{1}{2\sqrt{-h}} \partial_\alpha \left(
\sqrt{-h}h^{\alpha\beta} \partial_\beta \tilde{r} \right) 
= -\frac{1}{\tilde{r}_A} \left( 1-\frac{\dot{\tilde{r}}_A}{2H\tilde{r}_A} \right)
=-\frac{\tilde{r}_A}{2} 
\left( 2H^2+\dot{H} \right)$ and $h$ is the determinant of the non-spherical part of FLRW metric.
In contrast of Jacobson`s approach which temperature connects to local Rindler observers. Here, temperature could be measured by the Kodoma observer inside the apparent horizon. Kodama \cite{kodama} vector plays the role of the timelike Killing vector and if and only if the apparent horizon is trapping is positive. The Kodama temperature seems the natural candidate, because the Kodama vector is associated with a conserved current even in the absence of a timelike Killing vector (Kodama miracle).}
\begin{equation}\label{c5}
 T_{H}=\frac{1}{2\pi \tilde{r}_A} \left(
1-\frac{\dot{\tilde{r}}_A}{2H\tilde{r}_A} \right).
\end{equation}
Then, combining the equations (\ref{c3}) and (\ref{c5}) leads to
\begin{equation} \label{c6}
 T_{H} dS = 4\pi \tilde{r}_A^3 H \left(\rho_{t}+p_{t} \right)
dt
 -2\pi  \tilde{r}_A^2 \left(\rho_{t}+p_{t} \right) d\tilde{r}_A ~.
\end{equation}
We also have the total intrinsic energy as
\begin{eqnarray} \label{c7}
 dE = -4\pi \tilde{r}_A^3 H \left(\rho_{t}+p_{t} \right) dt
 +4\pi \tilde{r}_A^2 \rho_{t} d\tilde{r}_A ,
\end{eqnarray}
as well as   the work density  \cite{Hayward:1997jp, Hayward:1998ee, Cai:2006rs} as follows
\begin{equation} \label{c8}
 W \equiv -\frac{1}{2}  T^{\alpha\beta} h_{\alpha\beta} = \frac{1}{2} (\rho_{t}-p_{t}),
\end{equation}
where $T^{\alpha\beta}$ is the effective energy-momentum tensor of the EC
cosmos. Therefore, the unified first law of thermodynamics can be obtained in a straightforward manner by combining the equations (\ref{c6}), (\ref{c7}) and (\ref{c8})
as\begin{equation}\label{c9}
dE=- T_{H} dS+W dV.
\end{equation}
In addition, from the equation (\ref{c3}), we have 
\begin{equation}\label{jingel}
\dot{S} = - 2\pi \left[ \dot{H}/H^3  \right],
\end{equation}
for the surface entropy. Therefore, from the equations (\ref{c3}) and (\ref{jingel}), it is seen that if the null energy condition
holds,
i.e. $\rho_{t}+p_{t} \ge 0$, the surface entropy always increases in the expanding universe and we have  $\dot{H} \leq 0$.
\subsection{Generalized Second Law of Thermodynamics (GSL)}
In order to studying the generalized second law of thermodynamics, we consider
 the Gibbs equation 
\begin{equation}
 T_{H} dS_{b} = d\left( \rho_{t} V \right) +p_{t} dV =
V d\rho_{t} + \left( \rho_{t} +p_{t}\right) dV,
\end{equation} 
for the total matter content inside the bulk where we used the subscript $``b"$
to denote the entropy of inside of the bulk \cite{Karami:2012fu, Cai:2015emx,
Tian:2014ila}.
By combining the definition of the Hubble volume and equations (\ref{a}) and (\ref{b}), we obtain
 \begin{eqnarray}\label{c9.1}
T_{H} dS_{b} = \frac{\dot{H}}{H^4} (\dot{H}+H^2).
\end{eqnarray}   
Then, the total entropy can be divided into two parts, the total entropy inside the bulk $S_b$ and the part related to the surface $S$ as $ S_{t} \equiv S + S_{b}$.
By combining the modified Friedmann equations (\ref{a}) and (\ref{b}) and (\ref{jingel}), we have
\begin{eqnarray}\label{c10}
T_{H} \frac{dS_{t}}{dt}  = \frac{\dot{H}^2}{2H^4}. 
\end{eqnarray}
Consequently,   for an accelerating expanding universe with $H>0$, the generalized second law of thermodynamics always holds in the framework of the Einstein-Cartan cosmology.
\section{Covariant Entropy Conjecture and Emergent Universe scenario in Einstein-Cartan
theory}
  {In this section, we follow the approach of \cite{hamed}. We have the following condition  on the Padmanabhan's formula for an expanding Universe
\begin{equation}\label{a00}
\frac{d V}{d t}\geq0,
\end{equation}
which requires
\begin{equation}\label{c}
 N_{ sur}-N_{bulk}\geq0.
\end{equation}
where $N_{sur}$ and $N_{bulk}$  are given by the equations (\ref{googli}) and (\ref{Nbulk}). So, one can
rewrite the equation (\ref{c}) as follows
\begin{equation}\label{entropy11}
N_{spin} +N_{de}-N_{em}\leq N_{sur},
\end{equation}   
where using the $N_{sur}$  given by the equation (\ref{sur}), we have 
\begin{equation}\label{entropy00}
\frac{1}{4}\left (N_{spin} +N_{de}-N_{em}\right)\leq S.
\end{equation}} 
This relation  represents the existence of a lower bound for the entropy of a cosmological
system
in the framework of the emergent scenario. 
On the other hand,  the covariant entropy conjecture imposes an upper bound for the entropy of any thermodynamical system as \cite{Busso}
\begin{equation}\label{entropy0022}
S \leq \frac{A}{4},
\end{equation}
where $A$ is the area of the smallest sphere circumscribing the
system. Here, one may argue that it is not clear at all that how (\ref{entropy0022}) applies to the Einstein-Cartan theory, since the original derivation was for
general relativity. We refer the reader to the Appendix C, where we discussed
on the validity of (\ref{entropy0022}) in the context of the Einstein-Cartan theory. Then, for the Universe enclosed by the Hubble horizon $\tilde r_{H}$, we have $S\leq \pi \tilde r_{H}^2$. So, regarding the inequalities (\ref{entropy00})
and (\ref{entropy0022}), we find 
\begin{equation}\label{entropy33}
\frac{1}{4}\left (N_{spin} +N_{de}-N_{em}\right)\leq S\leq \pi \tilde r_{H}^2,
\end{equation} 
which gives a total restriction for the entropy of a cosmological system in the framework of the emergent scenario. In the absence of the dark energy component, the lower bound of the entropy may be takes negative value for the late times, due to
the vanishing behavior of the spin component (\ref{ns}). This is not physically acceptable and consequently the demand for the existence of the dark energy component is also seen here. One can also rewrite the inequality (\ref{entropy33})
as
\begin{equation}
\frac{4\pi V\sigma^{2}}{T}+ \frac{|\rho +3p|_{de}V}{2T} - \frac{(\rho +3p)_{em}V}{2T}\leq S\leq \pi \tilde r_{H}^2. 
\end{equation} 
Then, using the Hawking temperature  $T=H/2\pi$, the horizon radius $\tilde r_H =\frac{1}{H}$ and the Hubble volume $V=\frac{4}{3}\pi \tilde r_{H}^3$,
we arrive at the following inequality
regarding the upper bound
\begin{equation}\label{nn}
8\pi \sigma^{2}+ |\rho +3p|_{de}- (\rho +3p)_{em}\leq \frac{3}{4\pi}H^2,
\end{equation}
representing that the for such an expanding thermodynamical system, the total effective Komar energy never exceeds  the square of the expansion rate
with a factor of $\frac{3}{4\pi}$. Then, considering both of the covariant entropy
bound and the bound resulted from the emergent scenario  the evolution
of the density and pressure profiles in the Universe will be restricted as in  (\ref{nn}). The equality case occurs for the static state $H=0$, as for the pure de Sitter
universe,  and consequently we arrive at $ 8\pi \sigma^{2}+ |\rho +3p|_{de}- (\rho +3p)_{em}= 0$  indicating the balance between the effective repulsive and attractive
effects.
\section{Conclusion}
According to the Padmanabhan's emergent proposal, the accelerated expansion of the Universe
can be driven by the difference between the
surface degrees of freedom and the bulk degrees of freedom in a
region of space.  {The dynamical emergent equation is represented  by the relation $ dV/dt = N_{sur}-N_{bulk}$  where $N_{sur}$ and $N_{bulk}=N_{em}-N_{de}$ are  the degrees of freedom assigned
to the surface area and the matter-energy content inside the bulk, respectively
such that  the indexes $``em"$ and $``de"$  represent energy-momentum and dark energy, respectively. }In the present work,     spin degrees of freedom 
in the framework
of Einstein-Cartan (EC) theory are investigated.  In this regard, based on
the modification of Friedmann equations due to the spin-spin interactions, a correction term for the Padmanbhan's
relation including the number of degrees of freedom {related to this spin interactions is obtained as
$\Delta V /\Delta t=N_{sur}-N_{bulk}$ where $ N_{bulk}=  N_{em} -N_{spin} -N_{de}$
in which $N_{spin}$ is the corresponding degrees of freedom related
to the intrinsic spin of the matter content of the Universe. It is seen
that both of the spin  and dark energy sectors  contribute a negative
number of degrees of freedom.} Also, it is shown
that although the spin degrees of freedom can play an important role in the early stages of universe, but for the
late times the cosmological constant or dark energy term proposed by Padmanabhan is also required here to achieve the  holographic equipartition in this model. 
Moreover, the unified first law of thermodynamics for the Einstein-Cartan cosmos is obtained. It is shown that for an accelerating expanding universe, the generalized second law of thermodynamics always holds in the framework of this cosmological model.
  {Finally, by considering the covariant entropy conjecture and the bound resulted
from the emergent scenario, a total entropy bound is obtained. Using this bound,
it is shown that the for the universe as an expanding thermodynamical system, the total effective Komar energy never exceeds  the square of the expansion rate with a factor of $\frac{3}{4\pi}$.}
 \section*{Acknowledgment}
We would like to thank Amir Hadi Ziaie for useful comments.  This work   has been supported financially by Research Institute for Astronomy and Astrophysics of Maragha (RIAAM) under research project  No. 1/4717-43.
%
%
\section*{Appendix A}
For the expectation value of the $\tau^{\mu\nu}$ tensor given by the equation
(\ref{8}), we start with
\begin{eqnarray}\label{}
\tau^{\mu\nu}&=&8\pi G\{
-4\tau^{\mu\alpha}_{\hspace{.4cm}[\beta}\tau^{\nu\beta}_
{\hspace{.3cm}\alpha]}-2\tau^{\mu\alpha\beta}\tau^\nu_{\hspace{.3cm}\alpha\beta}
+\tau^{\alpha\beta\mu}\tau_{\alpha\beta}^{\hspace{.3cm}\nu}\nonumber\\
&&+\frac{1}{2}g^{\mu\nu}
\left(4\tau_{\lambda
\hspace{.1cm}[\beta}^{\hspace{.2cm}\alpha}\tau^{\lambda\beta}_{\hspace{.4cm}\alpha]}+
\tau^{\alpha\beta\lambda}\tau_{\alpha\beta\lambda}\right)\},
\end{eqnarray}
in which regarding the definitions in (\ref{FC}) and antisymmetric properties
with respect to $\alpha$ and $\beta$ indices, it reads as 
\begin{eqnarray}
<\tau^{\mu\nu}>&=&2\pi G\{-4u^{\mu}u^{\nu}<{S^{\alpha}}_{[\beta}{S^{\beta}}_
{\alpha]}>\nonumber\\
&&-2u^{\mu}u^{\nu}<S^{\alpha\beta}S_{\alpha\beta}>\nonumber\\
&&+u^{\mu}u^{\nu}<S^{\alpha\beta}S_{\alpha\beta}>\nonumber\\
&&+\frac{1}{2}g^{\mu\nu}
\left(4u_{\lambda}u^{\lambda}<{S^{\alpha}}_{[\beta}{S^{\beta}}_
{\alpha]}>\right)\nonumber\\
&&+\frac{1}{2}g^{\mu\nu}u_{\lambda}u^{\lambda}<S^{\alpha\beta}S_{\alpha\beta}>\},
\end{eqnarray}
which leads to 
\begin{eqnarray}
<\tau^{\mu\nu}>&=&2\pi G\{-2u^{\mu}u^{\nu}<\left({S^{\alpha}}_{\beta}{S^{\beta}}_
{\alpha}-{S^{\alpha}}_{\alpha}{S^{\beta}}_
{\beta}\right)>\nonumber\\
&&-2u^{\mu}u^{\nu}<S^{\alpha\beta}S_{\alpha\beta}>\nonumber\\
&&+u^{\mu}u^{\nu}<S^{\alpha\beta}S_{\alpha\beta}>\nonumber\\
&&+\frac{1}{2}g^{\mu\nu}
<2\left({S^{\alpha}}_{\beta}{S^{\beta}}_
{\alpha}-{S^{\alpha}}_{\alpha}{S^{\beta}}_
{\beta}\right)>\nonumber\\
&&+\frac{1}{2}g^{\mu\nu}u_{\lambda}u^{\lambda}<S^{\alpha\beta}S_{\alpha\beta}>\}.
\end{eqnarray}

Regarding that $S^{\mu\nu}$ is an antisymmetric tensor, then its trace vanishes
and consequently the second terms in the first and forth rows vanish. Then,
again due to anti-symmetry property of $S^{\mu\nu}$ and $\langle S_{\mu\nu}S^{\mu\nu}\rangle=2\sigma^2$, we arrive at 
\begin{equation}
<\tau^{\mu\nu}>=4\pi\sigma^2u^{\mu}u^{\nu}+2\pi G\sigma^2g^{\mu\nu}.
\end{equation}
One can use the same approach to find $<T_S^{\hspace{.2cm}\mu\nu}>=-8\pi G\sigma^2u^\mu u^\nu$ in (\ref{12}).
\section*{Appendix B}
One may argue about the negative contribution of the spin density in total
energy density (\ref{rev1})
and  violation of the energy conditions by the spin fluid.
One can find justification for this argument by looking at the Raychaudhuri equation. Indeed, a similar effect happens in the Raychaudhuri equation by the vorticity. This can be verified 
by the Raychaudhuri equation in the Einstein-Cartan universes  obtained in
 \cite{barrow} as
\begin{eqnarray}\label{raychad1}
\tilde \Theta^{\prime}&=&- \frac{1}{3}\tilde \Theta^2 -\frac{1}{2}\kappa(\rho+3p)-2\left(\tilde\sigma^2 -\tilde\omega^2   \right)
\nonumber\\
&& +\frac{1}{2}\kappa^2S^2 +...,
\end{eqnarray}
where  prime and tilde indicate purely Riemannian environments and $\tilde\sigma^2$,
$\tilde\omega^2$ and $S^2$ are the magnitudes of the shear, vorticity and spin tensors. This relation shows that the vorticity and spin/torsion
degrees of freedom have a rather similar nature acting against the attraction of gravity. Both of  the vorticity  and spin/torsion arise with
an opposite sign relative to the ordinary matter energy density in the Raychaudhuri
equation.

Also, one may argue about the absence of a real solution for $H$ through the relation (\ref{a})
at early time if  the spin 
fluid dominates to the usual perfect fluid. This issue is resolved in the
context of the non-singular cosmological models such as bouncing \cite{bp1,
bec1, bec2} or emergent cosmologies
\cite{ata}.  For these cosmologies at the bounce point, we have $H=0$ meaning
that the attracting effect of usual prefect fluid is balanced by the repulsion
effect of the spin/torsion fluid
leading to a bouncing solution for $H$. In this context, the time derivative of the Hubble parameter satisfies
the condition $\dot H>0$ at the bouncing point,  so that the universe possesses the
ability for transition from a contracting phase to an expanding one \cite{bp1,
bec1, bec2}. In such a scenario \cite{bec1}, as the universe contracts the total  density $\rho_{tot}$
increases like $1/t^2$, as usual, but then torsion kicks in and the maximal
density is reached. Then, as the universe continues to contract further, the density decreases, due to the negative contribution of spin fluid,  until it reaches zero
and a bounce occurs at the corresponding non-zero scale factor $a_0$. After the bounce, the density at first starts to increase with expansion, until the same maximum total energy density, i.e $\rho^{max}_{tot}$, is reached again. Then,  it  begins to decrease with the expansion according
to the usual behaviour as $\rho_{tot}\propto1/t^2$.  
Here, torsion induces a phantom period around the bouncing point such that
the total equation of state parameter, i.e $\omega_{tot}$, becomes infinitely negative at the bounce, because $p_{tot} <0$ is finite while $\rho_{tot} = 0$. After that, when torsion becomes sub-dominant, $\omega_{tot}$ goes to zero, as in usual cosmological history of the Universe.
\section*{Appendix C}
Here we discuss on the validity of (\ref{entropy0022}) in Einstein-Cartan
theory incorporating torsion field, regarding that  the original derivation was for
Einstein's GR.

The original derivation of the relation $S\leqslant A/4$ by Bousso was for the general relativity.
However, one may check the validity of this relation by checking its basic
requirements given in \cite{Busso}. In this Ref, we find the question "Given a two-dimensional surface B of area A, on
which
hypersurface H should we evaluate the entropy S?'' According to the Bousso's answer
to this question, we also find
the statement ``In order to construct a selection rule, let us briefly return to the limit in which
Bekenstein's bound applies. For a spherical surface around a Bekenstein system,
the enclosed entropy cannot be larger than the area. But the same surface is also a boundary of the infinite region on its outside. The entropy outside could clearly be anything. From this we learn that it is important to consider the entropy only on
hypersurfaces which are not outside the boundary''.
The terminology {\it ``outside''} is defined by Bousso as ``We start at B, and follow one of the
four families of orthogonal light-rays, as long as the cross-sectional area is decreasing or constant. When it becomes increasing, we must stop. This can be formulated
technically by demanding that the expansion 
of the orthogonal null congruence
must be non-positive, in the direction away from the surface B''. 
Therefore, here we just need to check that can we find such a null surface in the context of Einstein-Cartan
theory or not? To answer this question, we refer to  the Raychaudhuri equation in the Einstein-Cartan universes as obtained in the last section of \cite{barrow}. Following the authors of \cite{barrow}, the effect of spin fluid can be  highlighted further if we momentarily
consider the familiar general-relativistic scenario of purely gravitational "forces" acting on an
irrotational and shear-free perfect fluid with spin. Then, the equation (\ref{raychad1}) reduces to  
\begin{equation}\label{raychad}
\tilde \Theta^{\prime}=- \frac{1}{3}\tilde \Theta^2 -\frac{1}{2}\kappa(\rho+3p)
+\frac{1}{2}\kappa^2S^2 +\Lambda.
\end{equation}
Then, as discussed by the authors,  the spin term on the right-hand side of the above equation plays the role of an effective (positive)
cosmological constant (when $s = constant$), or that of a quintessence field (when $s = s(t)$). Thus, the spin effect in the Raychaudhuri equation of the
Einstein-Cartan universe by the Weyssenhoff fluid appears as a shift in the cosmological
constant of the Einstein general relativity theory. Then, the whole behaviour of
congruences of geodesic in Einstein-Cartan theory is the same as in GR, except
for a shift in the cosmological constant. Consequently, if we can find
the appropriate boundary surface in GR, then we are also able to define such a surface
in Einstein-Cartan theory. Therefore,  the application of entropy bound introduced by Bousso is also allowed in Einstein-Cartan theory.

%
%
%

\end{document}